\documentclass[sigconf,nonacm]{acmart}

\usepackage{graphicx}

\begin{document}

\title{A Comparative Study of DSL Code Generation: Fine-Tuning vs. Optimized Retrieval Augmentation}
\author{Nastaran Bassamzadeh\qquad Chhaya Methani\\Microsoft Corporation\\Redmond, USA}

\pagestyle{empty}


\begin{abstract}
  Natural Language to Code Generation has made significant progress in recent years with the advent of Large Language Models (LLMs). While generation for general-purpose languages like C, C++, and Python has improved significantly, LLMs struggle with custom function names in Domain Specific Languages or DSLs. This leads to higher hallucination rates and syntax errors, specially for DSLs having a high number of custom function names. Additionally, constant updates to function names add to the challenge as LLMs need to stay up-to-date. In this paper, we present optimizations for using Retrieval Augmented Generation (or RAG) with LLMs for DSL generation along with an ablation study comparing these strategies. We generated a train as well as test dataset with a DSL to represent automation tasks across roughly 700 APIs in public domain. We used the training dataset to fine-tune a Codex model for this DSL. Our results showed that the fine-tuned model scored the best on code similarity metric. With our RAG optimizations, we achieved parity for similarity metric. The compilation rate, however, showed that both the models still got the syntax wrong many times, with RAG-based method being 2 pts better. Conversely, hallucination rate for RAG model lagged by 1 pt for API names and by 2 pts for API parameter keys. We conclude that an optimized RAG model can match the quality of fine-tuned models and offer advantages for new, unseen APIs.
\end{abstract}

\keywords{Code Generation, NL2CodeGen, DSL, NL2DSL, RAG, Fine-Tuning}

\maketitle

\section{Introduction} \label{intro}
There has been significant progress made in improving and quantifying the quality of Natural Language to Code Generation or NL2Code (\cite{Chen21CodeGenEval}, \cite{Nguyen22Eval}, \cite{Feng20CodeBERT}, \cite{Chen21CodeGenEval}). Recent improvements in models for general-purpose languages like Python, C++ and Java can be attributed to larger LLMs (\cite{ChatGPT22}, \cite{GPT423}) and the availability of pre-trained open-source models (\cite{Feng20CodeBERT}, \cite{CodeLlama23}, \cite{abdin2024phi3}, \cite{Codestral24}) advancing the state-of-the-art. However, there hasn’t been a focus on improving quality of Natural Language to Domain Specific Languages or NL2DSL which a lot of enterprise applications rely on. 

Domain Specific Languages (or DSLs) are custom Computer Languages designed and optimized for specific applications. Examples of DSLs include SQL and industry-specific languages for formalizing API calls, often using formats like JSON or YAML to represent API sequences. In this paper, we focus on the task of generating a DSL used for authoring high-level automation workflows across thousands of web-scale APIs. These workflows support a variety of customer scenarios like invoice processing, sales lead integration with forms/emails etc. The automation DSL represents API names as functions and codifies a sequence of API calls along with conditional logic over the invocation of APIs. We constrained the length of sequence to 5 APIs and hope to explore longer sequences in future work. An example of the DSL is shown in Figure \ref{fig:system_arch}.

Existing code generation methods are hard to adapt for this scenario due to the frequent hallucinations and syntax errors. This is largely due to the custom names, massive size and diversity of APIs in public as well private domain along with the ever-changing API landscape. Current NL2Code methods mainly use fine-tuning and do not focus on strategies for improving grounding LLMs to include new APIs.

In this paper, we outline an end to end system architecture for NL2DSL generation with high response rate using selective improvements to RAG techniques (\cite{Liu23ChatGPTPrompts}, \cite{Poesia22TST}) using OpenAI models. We fine-tuned a Codex model for NL2DSL and show a comparative analysis of the impact of the approaches used to optimize RAG. 

Along with metaprompt tuning for RAG, we also included additional grounding context in the form of API Function Definitions, like the approach used for Tool Selection (\cite{Schick23Toolformer},\cite{Shen23HuggingGPT}, \cite{liang2023taskmatrixai}, \cite{Patil23Gorilla}). This is motivated by the similarities between the code generation and task orchestration scenarios discussed in more detail in Section \ref{relatedwork}.

The remainder of this study is structured as follows. In Section \ref{relatedwork}, we present the NL2DSL problem formulation along with literature review. The focus is on comparing differences between Tool Selection of APIs as a framework compared to Code Generation over a set of APIs. This will help define the scope of the experiments in this study. Section \ref{methodology} lays out and describes the optimizations we made to RAG as discussed above along with the benchmark Fine-Tuned model. Section \ref{expresults} discusses Data Generation, Metric definition and Section \ref{results} shares our results and discussion followed by Conclusion and Future Work in Section \ref{conclusion}.

\section{Related Work} \label{relatedwork}

\subsection{Code Generation or Program Synthesis}

Program Synthesis is a hard research problem (\cite{jain22jigsaw}, \cite{devlin2017robustfill}, \cite{Feng20CodeBERT},\cite{Li_2022}, \cite{xu2021inide}). It has gained significant interest with many open-source models focusing on general programming languages since the release of Github Copilot (\cite{Chen21CodeGenEval}). These models include Code Llama \cite{CodeLlama23}, StarCoder \cite{li2023starcoder},  Codestral \cite{Codestral24}, Phi-3 \cite{abdin2024phi3} and more.  Many of these advancements have been achieved through pre-training language models for code generation with a focus on improving datasets((\cite{CodeLlama23}, \cite{abdin2024phi3})). However, for domain adaptation, \textbf{instruction fine-tuning} on top of a base model remains a popular approach (\cite{Chen21CodeGenEval}, \cite{gao2023pal}, \cite{lewkowycz2022solving}, \cite{Patil23Gorilla}). 

Prompting LLMs is an alternative technique for code generation (\cite{Liu23ChatGPTPrompts}, \cite{White23ChatGPTPrompts}, \cite{wei2023chainofthought}, \cite{kojima2023large}). Poesia et al. (\cite{Poesia22TST}) focused on improving response quality through grounding techniques. They fine-tuned a Sentence BERT model by changing the loss function to incorporate predicting similarity of the generated target programs. With this adapted similarity metric, better few shots are selected dynamically.

\subsection{Reasoning and Tool Integration}
When it comes to modeling the problem of selecting a sequence of API calls, we need to consider formulating it as a \textbf{planning} or \textbf{reasoning} task. LLMs show remarkable reasoning capability, however, they also have limitations when it comes to staying up-to-date with recent knowledge, performing mathematical calculations etc. A popular way to overcome this has been granting the LLMs access to external tools. This framework gained significant popularity with OpenAI Code Interpreter's success (\cite{CodeInter23}). 

External Tool Integration has been studied since with a focus on including specific tools such as web search (\cite{Schick23Toolformer}), python code interpreters (\cite{gao2023pal}, \cite{CodeInter23}), adding calculators (\cite{Parisi2022Talm} \cite{gao2023pal}) and so on. Expanding the tool set to a generic list of tools has been explored (\cite{Schick23Toolformer}, \cite{Patil23Gorilla}), but it remains limited and often predicts single tools instead of sequences needed for most enterprise scenarios. Tool Use has mostly been explored in the context of generating more accurate text outputs for Q\&A tasks with the help of external tools(\cite{Schick23Toolformer}, \cite{Parisi2022Talm}).

There is an increase in focus on incorporating LLM's code generation capabilities to reasoning and task orchestration, this is an area of active research (\cite{gao2023pal}, \cite{liang2023taskmatrixai}, \cite{Patil23Gorilla}). However, most of the research either limits the tools to a set of small well-documented APIs ( (\cite{gao2023pal},\cite{liang2023taskmatrixai}), or limited their scope to predicting a single output API (\cite{Patil23Gorilla}, \cite{Schick23Toolformer}). 

Posing the reasoning or orchestration task as a code generation problem is similar to the API sequence generation scenario highlighted in this paper. Representing a plan as a DSL, as discussed in Section \ref{intro}, aligns with our goal of generating DSL for workflow automation. Improving the quality of Natural Language to DSL generation, is thus beneficial for both reasoning and plan generation.

\subsection{Contributions}
In the previous section, we discussed formulating Task Orchestration as a Code Generation problem since it can be represented as yet another DSL. NL2DSL generation suffers from the hallucination and quality issues we discussed in \ref{intro}. Few studies address the challenges of end-to-end DSL generation, specifically over a large set of custom APIs.  

\begin{figure*}
  \centering
  \includegraphics[width=\textwidth]{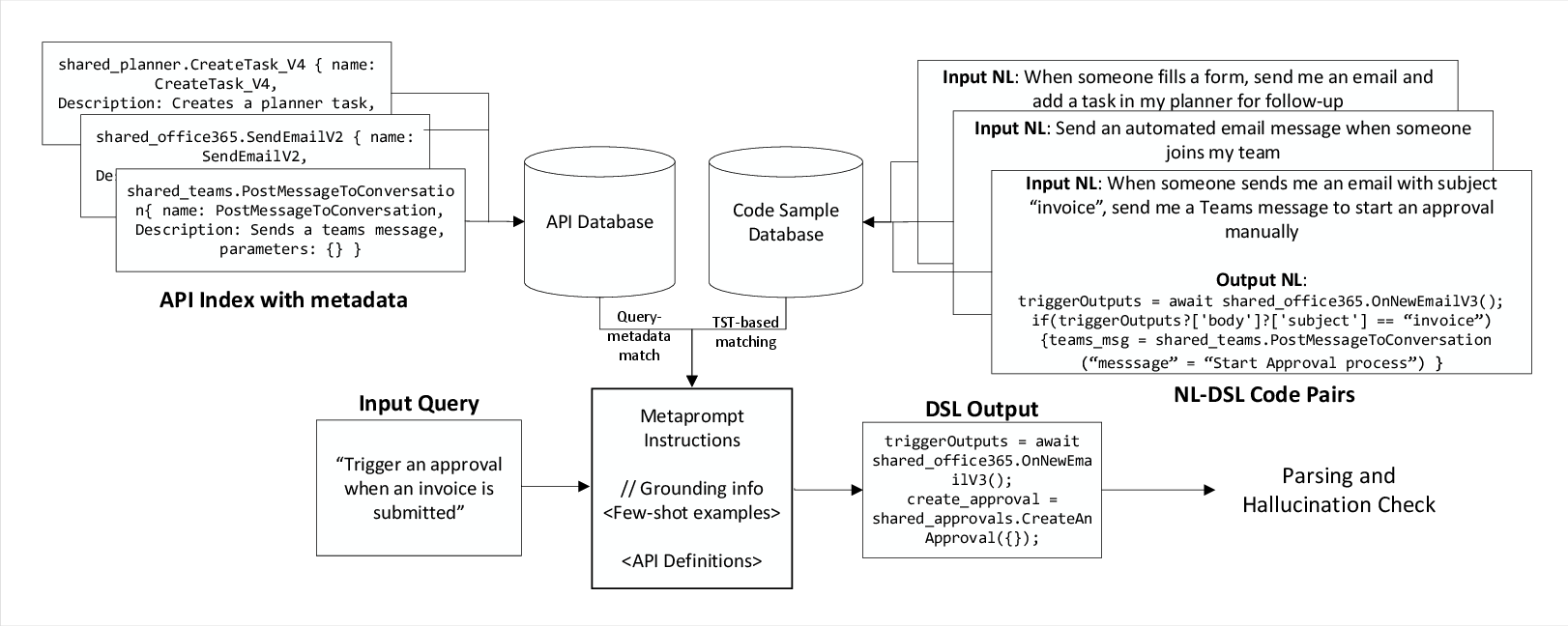}
  \caption{System Architecture to show e2e working of \&
    our DSL generation methodology using RAG. TST based semantic mapping \&
    retrieves the relevant code snippet as shown. This helps get the right syntax. However, \&  it gets the correct function name for approval from the API metadata\&}
  \Description{System Architecture along with NL and DSL pairs from our dataset. Note, function names are indicative to show API functionality for this illustration.}
  \label{fig:system_arch}
\end{figure*}

This paper presents improvements to known RAG techniques focusing on improving DSL generation quality for enterprise settings. Our DSL expands API or tool selection to a sequences of 5-6 API calls, also referred to as chain of tools, which is a first to the best of our knowledge. We also consider the real-world scenarios of adding conditional logic with API calls as shown with an example in Figure \ref{fig:system_arch}. Our contribution is outlining an end-to-end system as well as presenting an ablation study for NL2DSL generation. 

We merged prompting and grounding approaches from code generation (\cite{Poesia22TST},\cite{Liu23ChatGPTPrompts},\cite{White23ChatGPTPrompts}) and added API metadata as used in task orchestration area (\cite{gao2023pal}, \cite{liang2023taskmatrixai}) and studied their impact on reducing hallucination rate. We created a test set having 1000 NL-DSL pairs spanning over a set of approx. 700 API calls or functions using principles of synthetic dataset generation (similar to \cite{honovich2022unnatural} and \cite{schick-schutze-2021-generating}) and used manual approval to validate test set quality. Our fine-tuned DSL model is trained on a larger synthetic NL-DSL dataset (details in Section \ref{ss:fine-tuning}).  

\section{Methodology} \label{methodology}
In this section, we first provide an overview of the approaches used in our experiments. In the following sub-sections, we will delve deeper in the details of each of the approaches.

Details of fine-tuning are shared in Section \ref{ss:fine-tuning}. Fine-Tuning a base model, specifically, instruction fine-tuning is a preferred approach for domain adaptation. It's limitations include inability to include newly added APIs on an ongoing basis, as well as the resource intensive data collection process for infrequently used APIs or the tail set. 

We used RAG based approaches to overcome these limitations, and focused on improving grounding techniques for DSL generation (Details in Section \ref{ss:RAG}). We used dynamically generated few-shot examples approach (\cite{Poesia22TST}), and augmented it with API function definitions similar to the way it is used for Tool Selection (\cite{Patil23Gorilla}, \cite{Shen23HuggingGPT}). These few-shots were selected from an expansive pool of synthetic NL-DSL pairs, empirically having 100s of variations of usage for each API (Section \ref{ss:dataset}). 

For computing semantic similarity of the few-shots with the input user query (Section \ref{ss:RAG}), we fine-tuned a BERT model as highlighted in \cite{Poesia22TST} with a modified loss function for predicting target DSL similarity. For selecting the API metadata for grounding (Section \ref{ss:tool_defs}), we created an index over API Function Definitions. We also tried metaprompt tuning, but limit the focus of this study to improving grounding techniques with a combination of dynamically selected few-shot samples as well as API metadata or tool description. 

We share the details of each approach and variation below.

\subsection{Fine-Tuned NL2DSL Generation Model} \label{ss:fine-tuning}
We took the Codex base model from OpenAI due to it's pre-training with code samples and used LoRA-based fine-tuning approach. The training set consists of NL-DSL pairs, NL refers to the user query and the DSL represents the workflow that the user is looking to automate. We used <START> and <END> token to indicate the end of code generation to the model.  
The training set consists of a pool of 67k samples in the form of (prompt, flow) tuples generated synthetically ( details in Section \ref{ss:dataset}, and examples of NL-DSL are shared in Figure \ref{fig:system_arch} and Appendix \ref{sec:appendix}).

We ran many iterations on this model to improve performance on the test set, specifically for the body and tail connectors, and went through multiple rounds of data augmentation. We found that predicting the parameter keys was very challenging with the fine-tuned model due to limitation of data generation. Even with synthetic models, it was hard to scale the NL-DSL sample variety needed for improving quality of parameters.  

\subsection{Grounding with dynamically selected few-shots} \label{ss:RAG}
We tried two types of grounding information for RAG based DSL generation as described below. There are some variations of each technique described in the paragraph below as well. For each technique, we selected $5$ and $20$ shots dynamically, and saw performance impact driven by the approach used for sample selection.

\subsubsection{Pre-Trained Model}
The first approach is using a vanilla Per-Trained model for determining the semantic similarity of NL-DSL samples based on the NL query. We computed the embeddings of NL queries using a Distil-RoBERTa Pre-Trained model. We created a Faiss Index (\cite{Faiss2024}) for these embeddings to help with search over the dense embedding space.

\subsubsection{TST based BERT Fine-tuning}
In this approach, we fine-tuned the pre-trained model to improve the retrieval accuracy of the few-shots. This is similar to the approach used by Poesia et al. in \cite{Poesia22TST}. They show that if we fine-tune the pre-trained BERT model with a modified loss function to consider the similarity between the target DSL for each NL-DSL pair, the retrieved examples will have a higher quality and finally lead to better generation with LLM.

To get positive and negative samples for fine-tuning, we compared cosine similarity between all pairs of Natural Language queries in our dataset (Dataset shared in Section \ref{ss:dataset}). We used a Pre-Trained Tansformer model to generate embeddings for the purpose of similarity computation. A pair of tuples is considered a positive sample if the similarity between their corresponding NL prompts is greater than 0.7 and negative otherwise. We generated 100k pairs this way and leveraged them as training data for our fine-tuning experiment. 

The loss function used by TST (Equation \ref{MSE_loss} from \cite{Poesia22TST}) is minimizing the Mean-Squared Error between the vanilla loss functions comparing the utterances ($u_i,u_j$) and the target programs ($p_i,p_j$). Program similarity is denoted by $S$. They used AST to compute program similarity, however, we used a Jaccard score over lists of API function names to be consistent with our metrics definition (Section \ref{expresults}). 

\begin{equation} \label{MSE_loss}
L_{TST}(\theta):= E_{i,j ~ D} [f_{\theta}(u_i, u_j) - S (P_i,p_j)]^2
\end{equation}

\subsection{Grounding with API Metadata} \label{ss:tool_defs}
In addition to few-shots, we appended the API metadata in the metaprompt. This metadata includes Function Description along with the parameter keys and their description (See an example API Function Definition shared in Appendix \ref{sec:appendix}). We followed the below two approaches for selecting the metadata to be added. 

\subsubsection{API Function Definitions for Few Shots}
For the few-shots samples selected using the methods described above, we extracted the metadata for \textbf{each of} the functions present in those samples. This means that for the $n$ few-shot samples dynamically added to the metaprompt, we iterated over all the API function names in each of these flows and added their function definitions to the metaprompt. 

We also modified the metaprompt to add instructions on how to use the Function Definitions. We want to explore how adding the metadata explaining the purpose of each function in the few-shot examples impacts LLM's understanding of the task and map to user request.

\subsubsection{Semantic Function Definitions}
Another approach for selecting the function definitions to be added to the metaprompt is to retrieve the semantically similar functions from a vector database created with API metadata. This approach is similar to the one followed by LlamaIndex (\cite{LlamaIndex}) We created an index of all API definitions and retrieved the semantically similar functions by using the input NL query to search the index. Please note that this is different from the faiss index created for few-shot samples in Section \ref{ss:RAG}.

We call this approach \textbf{Semantic Function Definition (SFD)} and will compare it with the \textbf{Regular FDs} described above. This approach can be specifically useful for tail-ish prompts where no few-shots might be retrieved. This helps us integrate the newly released web APIs in our DSL Generation framework making our approach scalable to the changing API landscape.

\section{Experiment Design and Metrics Definition} \label{expresults}
In this section, we outline the process of Dataset Generation and introduce the metrics we used for estimating the code quality. We then describe the experiments. Results and Discussion follows in the next section. We have used Azure AML pipelines to run our experiments. The GPT-4 (with 16k token limit) model is used as the LLM model. The metaprompt is kept consistent between experiments for the purpose of the ablation study.

\subsection{Dataset Generation} \label{ss:dataset}
We generated a total of 67k samples in the form of (prompt, flow) pairs from workflows created by users. We had many samples of workflow automations created by users across a large set of APIs. We sampled the automations containing $700$ publicly available APIs and synthetically generated the corresponding Natural Language prompts using GPT-4. For creating these NL descriptions for the workflows, we also provided API Function definitions to the metadata. This ensured the language of the description captured the functioanlity of the API.

A subset of these synthetic samples were validated by human judges. We used these checks to improve the metaprompt used for synthetic data generation. For creating a test set, we used the same  process with most of the test set evaluated by human judges to ensure quality. We followed the same distribution of APIs from users, to ensure that our metrics are not biased. The test data set consists of $1000$ samples that are verified by human judges.

\subsection{DSL Generation Quality Metrics}
We defined 3 key metrics to focus on code generation quality as well as syntactic accuracy and hallucination rate. We have a compiler to test the syntax and validate the functions against a database of API names as well as parameter keys.

\subsubsection{Average Similarity} 
Average Similarity measures the aggregated similarity between predicted flow and the ground truth flow. The average similarity between two flows is defined using the Longest Common Subsequence match (LCSS) metric. Each flow is reduced to a list of API call sequences and then the LCSS is computed. The final metric is reported as an average over all test samples. Hallucination and Parser failures lead to the sample being discarded and is assigned a similarity score of 0.

\begin{equation}
    \textrm{Similarity} = \frac{\mathrm{LCSS} (A, B)} {max (|\mathrm{Actions}_A|, |\mathrm{Actions}_B|)} 
\end{equation}
where $|\textrm{Actions}_A|$ is the number of actions in flow $A$ and $|\textrm{Actions}_B|$ is the number of actions in flow $B$.
    
Please note that we are not using the commonly used AST metric for computing code similarity. AST drills down to compare similarity performance for parameters as well. As we wanted to focus on the problem of improving function name retrieval as well as it's sequence, we chose to define the metric in this manner. 


\begin{table*}[ht]
  \caption{Compare impact of selecting \textbf{5 vs 20 few shot} samples for both TST vs. Pre-trained Model without adding API function definitions using GPT-4. All results are shown as $\Delta$ improvements compared to the baseline. The baseline uses Pre-Trained Transformer Model with 5 few-shot samples. For Avg. similarity, higher is better, and for the rest of metrics capturing failure rates, lower is better.}
  \label{tab:fewshotcomparisonmodel5n20}
  \begin{tabular}{lccccc}
    \toprule
    Model &  Num of Few-Shots&Avg. Similarity & \%non-parsed flows & \%made-up API names & \%made-up parameters\\
    \texttt{Pre-trained Model w\/o FD} &  20&\boldmath{$+0.03$}& \boldmath{$-3.37$}& $-7.34$& \boldmath{$-15.17$}\\
    \texttt{TST w\/o FD} &  5&$+0.02$& $-0.61$& $-3.53$& $-1.04$\\
    \texttt{TST w\/o FD} &  20&\boldmath{$+0.03$}& $-2.85$& \boldmath{$-8.49$}& $-14.58$\\
   
    \bottomrule
  \end{tabular}
\end{table*}

\begin{table*}
  \caption{Impact of selecting \textbf{5 few shot} samples using TST vs. Pre-trained Model with and without API Function Definitions using GPT4 model. All results are shown as $\Delta$ improvements compared to the baseline. The baseline uses Pre-Trained Transformer Model without API Function Definitions. For Avg. similarity, higher is better, and for the rest of metrics capturing failure rates, lower is better.} 
  \label{tab:fewshotselectionmodel}
  \begin{tabular}{lcccc}
    \toprule
    Model & Avg. Similarity & \%Unparsed flows & \%made-up API names & \%made-up API parameters\\
    \texttt{Pre-trained Model + FD} & $0$ & $+2.75$ & $-4.3$ & \boldmath{$-20.16$} \\
    \texttt{TST w\/o FD} & \boldmath{$+0.02$} & \boldmath{$-0.61$} & $-3.53$ & $-1.04$ \\
    \texttt{TST + FD} & \boldmath{$+0.02$} & $+0.68$ & \boldmath{$-6.29$} & $-19.99$ \\
    \bottomrule
  \end{tabular}
\end{table*}

\begin{table*}
  \caption{Impact of selecting \textbf{20 few shot} samples using TST vs. Pre-trained Model with and without function definitions using GPT4 model. All results are shown as $\Delta$ improvements compared to the baseline. The baseline uses Pre-Trained Transformer Model without API Function Definitions. For Avg. similarity, higher is better, and for the rest of metrics capturing failure rates, lower is better.}
  \label{tab:fewshotselectionmodel20}
  \begin{tabular}{lcccc}
    \toprule
    Model & Avg. Similarity & \%Unparsed flows & \%made-up API names & \%made-up API parameters\\
    \texttt{Pre-trained Model + FD} & $-0.01$ & $+2.29$ & $-2.17$ & $-6.93$\\
    \texttt{TST w\/o FD} & $0$ & \boldmath{$+0.52$} & $-1.15$ & $+0.52$\\
    \texttt{TST + FD} & \boldmath{$+0.02$} & $+0.83$ & \boldmath{$-2.7$} & \boldmath{$-7.06$}\\
    \bottomrule
  \end{tabular}
\end{table*}
\subsubsection{Unparsed rate}
This metric captures the rate of syntactic errors. A flow that cannot be parsed by the parser is considered not usable for the purpose of similarity metric computation. Unparsed rate is computed as follow:

\begin{equation}
    \%\mathrm{unparsed \ flows} = \frac{|\mathrm{Flows}_\mathrm{unparsed}|}{|\mathrm{Flows}_\mathrm{total}|}
\end{equation}
where, $|\mathrm{Flows}_\mathrm{unparsed}|$ is the number of flows that were not parsed and $|\mathrm{Flows}_\mathrm{total}|$is the total number of flows in the sample set.
        
\subsubsection{Hallucination rate}
This metric captures the rate of made-up APIs (or function names) and made-up parameter keys in the generated code. Predicting a flow with a hallucinated API name is counted as a failure and leads to the code being considered invalid. 

We compute this by counting the number of flows that have at least one hallucinated function name and divide it by the total number of flows in the sample set.

\begin{equation}
    \%\mathrm{made-up \ APIs} = \frac{|\mathrm{Flows}_h|}{|\mathrm{Flows}_\mathrm{parsed}|} * 100
\end{equation}

where $|\mathrm{Flows}_h|$ is the number of flows with hallucinated API names and $|\mathrm{Flows}_\mathrm{parsed}|$ is the number of flows that were parsed correctly.

Similarly, we compute the rate at which parameters were not parsed. Failure to parse parameters does not result in the flow being discounted from average similarity computation. However, it shows up as run-time errors. Fixing these run-time errors is beyond the scope of this paper.
\begin{equation}
    \%\mathrm{made-up \ parameters} = \frac{|\mathrm{Flows}_{hp}|}{|\mathrm{Flows}_\mathrm{parsed}|} * 100
\end{equation}

where, $|\mathrm{Flows}_{hp}|$ is the number of flows with hallucinated parameter key names and $|\mathrm{Flows}_\mathrm{parsed}|$ is the number of flows that were parsed correctly.

\section{Results} \label{results}
In this section, we present the results of the above approaches on a test set of 1000 NL-DSL pairs. These samples, while generated synthetically, have been evaluated by human judges for quality. They were also sampled to represent the distribution of APIs in actual product usage.

We compare the impact of each ablation in sections below.

\subsection{Impact of number of few-shots on RAG performance}
We compare the impact of number of code samples added to the meta prompt with two different settings i.e. 5 few-shots vs 20 few-shots. We measured the results for both Pre-Trained model as well as TST model. Results are shared in Table \ref{tab:fewshotcomparisonmodel5n20} and show the $\Delta$ change compared to that Baseline model. The baseline setting here is Pre-Trained model with 5 few-shots.

Looking at row 1 and comparing rows 2 and 3 with respect to the baseline , we can see that adding more few-shots improves the performance of both the Pre-Trained as well as the TST model on all  metrics. The gain is particularly pronounced for reducing the number of made-up API names as well as reducing the number of made-up API parameter keys. We saw the gain plateau beyond this,  and we intend to run more experiments in the future to study this effect better.

\begin{table*}
  \caption{Impact of adding API or tool related metadata on performance (with GPT-4 model and 20 few shots). FD refers to including only metadata for APIs present in few-shots. SFD refers to extracting APIs similar to the input query (Refer to Section \ref{methodology}) for details.   The baseline uses fine-tuned Codex model. For Avg. similarity, higher value is better, and for the rest of metrics capturing failure rates, lower is better. }
  \label{tab:apiselectionmodel}
  \begin{tabular}{lcccc}
    \toprule
    Model & Avg. Similarity & \%Unparsed flows& \%made-up API names& \%made-up API parameters\\
    \texttt{TST + FD} & \boldmath{$0$}& \boldmath{$-5.3$}& $+1.7$& \boldmath{$+1.11$}\\
    \texttt{TST + SFD} & $-0.01$& $-1.43$& $+1.21$& $+6.76$\\
    \texttt{TST + FD + SFD} &\boldmath{$0$} & $-2.74$& \boldmath{$+0.94$}& $+2.03$\\
    \bottomrule
  \end{tabular}
\end{table*}

\subsection{TST vs Pre-trained Model}
Comparing the rows in Table \ref{tab:fewshotcomparisonmodel5n20}, both Pre-Trained and TST with 20 samples look comparable for computing the Average Similarity but have slight variations in Unparsed flow rate as well as Hallucinations rates. TST model performs better in reducing the $\%$ made-up API names, while the Pre-trained model has a slight edge in the other two metrics. 

So, we additionally look at the impact of including API Function Definitions to both the models (see Table \ref{tab:fewshotselectionmodel}). Here, we have used GPT4 model with 5 few shots. The results are represented as $\Delta$ changes compared to the Baseline setting i.e. using the Pre-Trained model to choose $5$ few-shot NL-DSL code samples. TST with FD setting performs overall better than all other options with values close to the best in every metric.

We see a similar trend  in Table \ref{tab:fewshotselectionmodel20}  where we captured the results for $20$ few-shots. This leads us to conclude that the presence of few-shot examples is supported by adding the API functions definitions of these functions (as described in Section \ref{methodology}). The addition predominantly helps reducing the hallucination rate for API names and parameters, which improves the overall response rate of NL2DSL generation. 

\subsection{Function Definition vs Semantic Function Definitions} 
As the next step, we will compare the impact of Semantic Function Definitions (SFD) vs adding the API Function Definitions for selected examples only. We used a Fine-Tuned model as baseline for this experiment. Based on the insights from the previous step, we used 20 few-shots for TST along with including FDs.  All results in Table \ref{tab:apiselectionmodel} are shown as $\Delta$ improvements compared to the baseline. 

Looking at metrics in columns for $\%$ made-up API names and $\%$ made-up parameter keys, we see that the hallucination rate is in general increasing for RAG based approach. However, we need to keep in mind that a fine-tuned model on the function names is hard to beat as it has been trained on 67,000 samples compared to only 20 few-shots that have been added to the RAG model.  

Within the RAG approaches, comparing rows 1 and 2 ("TST + FD" vs "TST + SFD"), SFD in general results in a slight drop in average similarity and an increase in the Unparse rate as well as hallucination rate for parameter keys.  This indicates that the approach to simply add semantically similar API metadata for a query is not useful for DSL generation. We get better similarity, as well as reduced Hallucination Rate when we include the API Function Definitions for the samples selected by TST (as shown in Row 1).

The addition of Semantically matching Function Definitions tends to reduce the hallucination of API names indicating that it could have potential of adding FDs that are not a part of the code sample set. This could have implications for improving the performance for newly added APIs in the public cloud, that will help keep the performance of the system updated. We will explore this topic in a future study.

\section{Conclusion} \label{conclusion}
Concluding from the ablations study shared in Section \ref{results}, we see that the role of dynamically selected few-shot samples is very important in making RAG useful for syntactically correct generation of DSL as well as improving code similarity ((Table \ref{tab:apiselectionmodel})).  

Fine-Tuning still outperforms the RAG based model in terms of lower hallucinations (see Table \ref{tab:apiselectionmodel} where fine-tuned model is the baseline). However, the parsing errors are more common in the fine-tuned model. This could be due to the fact that few shot examples have been successfully teaching the correct syntax to the LLM model. It is, however, surprising that the syntax correctness for RAG is better than that of the fine-tuned model which was trained on a much larger sample set. 

It is also interesting to note that this benefit does not transfer to hallucinated API names and their parameters keys where the fine-tuned model holds the advantage. The increase of 6.76 pts in hallucination rate for parameters due to adding Semantic Function definitions indicates that adding too many API descriptions can confuse rather than help the LLM (Table \ref{tab:apiselectionmodel}). It also signifies the higher impact of the few shot samples for the scenario of DSL Generation or API selection compared to simply providing the API description. This learning can be used to inform the Tool Selection or orchestration scenario. Providing high quality examples of sample orchestration will reduce the failure rate more. 

Overall, we were able to significantly improve the performance of RAG for DSL generation, with hallucination rate for API names dropping by 6.29 pts. and that of parameter keys dropped by approx. 20 pts (see Table \ref{tab:fewshotselectionmodel}). The performance of RAG is now comparable to that of fine-tuned model (see Avg. Similarity in Table \ref{tab:apiselectionmodel}), with the potential to bootstrap quickly. Optimized RAG can also allow extending the benefits of metaprompt tuning to include  unseen APIs, reducing the need to fine-tune the model frequently. This will be the focus of our future work. 

\bibliographystyle{ACM-Reference-Format}
\bibliography{sample-base}

\appendix

\section{Appendix} \label{sec:appendix}

\subsection{Sample with computed Average similarity} \label{ssec:simialrity}

Sample showing how flow similarity is computed for two flows Flow A and Flow B.

\begin{verbatim}

Query = "Post a message in the channel of teams, 
when a new form is created in the forms"

Ground Truth = "triggerOutputs =
await shared\_microsoftforms.CreateFormWebhook({}); 
outputs_shared_teams_PostMessageToConversation = 
shared_teams.PostMessageToConversation(
{ \"poster\": \"User\" });"    

prediction: "triggerOutputs = 
await shared_microsoftforms.CreateFormWebhook({});
outputs_Get_my_profile_V2 =  shared_office365users.
MyProfile_V2({}); outputs_shared_teams_PostMessage
= shared_teams.PostMessageToConversation(
{\"poster\": \"User\",\"location\": \"Channel\"});"

API Functions list in ground_truth  = 
[shared_microsoftforms.CreateFormWebhook, 
shared_teams.PostMessageToConversation]


API function list in model generation = 
[shared_microsoftforms.CreateFormWebhook, 
shared_office365users.MyProfile_V2,
shared_teams.PostMessageToConversation]

Similarity Score = 2/3 = 0.666

Since the functions shared_microsoftforms.
CreateFormWebhook and shared_teams.
PostMessageToConversation are found 
in the ground truth.
\end{verbatim}

\subsection{An example of API metdata}
We share a sample of API metadata to highlight the details included in the API description provided to the metaprompt.

\begin{verbatim}

"shared_outlook.SendEmailV2": {
    "FunctionName": "shared_outlook.SendEmailV2",
    "Description": "This operation sends an email message.",
    "IsInTrainingSet": false,
    "DisplayName": "Send an email (V2)",
        "ParametersInfo": [
            {
                "Key": "emailMessage/To",
                "Type": "String",
                "Summary": "To",
                "Format": "email",
                "Description": "Specify email addresses 
                    separated by semicolons like 
                    someone@contoso.com"
            }, ….
                        ],
        "ResponseSchema": [],
        "IsTrigger": false
    }

\end{verbatim}

\end{document}